\documentclass[byrevtex,twocolumn,prb,showpacs]{revtex4}
\usepackage{epsfig}
\usepackage{amsmath}
\usepackage{amssymb}
\usepackage{times}
\usepackage{graphicx}
\begin{document}

\title{Spin current diode based on an electron waveguide with spin-orbit
interaction}
\author{Feng Zhai}
\email{fengzhai@dlut.edu.cn} \affiliation{School of Physics and
Optoelectronic Technology and College of Advanced Science and
Technology, Dalian University of Technology, Dalian 116024,
People's Republic of China}
\author{Kai Chang}
\email{kchang@red.semi.ac.cn} \affiliation{NLSM, Institute of
Semiconductors, Chinese Academy of Sciences, P.O. Box 912, Beijing
100083, People's Republic of China}
\author{H.~Q. Xu}
\email{Hongqi.Xu@ftf.lth.se}
\affiliation{Division of Solid State Physics, Lund University, Box 118, S-22100 Lund,
Sweden}

\begin{abstract}
We propose a spin current diode which can work even in a small
applied bias condition (the linear-response regime). The
prototypal device consists of a hornlike electron waveguide with
Rashba spin-orbit interaction, which is connected to two leads
with different widths. It is demonstrated that when electrons are
incident from the narrow lead, the generated spin conductance
fluctuates around a constant value in a wide range of incident
energy. When the transport direction is reversed, the spin
conductance is suppressed strongly. Such a remarkable difference
arises from spin-flipped transitions caused by the spin-orbit
interaction.
\end{abstract}

\maketitle

The spin of carriers has been exploited in recent years to develop
solid-state devices combining the standard microelectronics with
spin-dependent effects.\cite{Spintronics,Spintronics1} The operation of
spin-based electronic circuits requires the electrical generation of excess
spin in nonmagnetic materials. To this end, various spin injection methods
and spin filters have been explored. For a spin filtering device it would be
very desirable that the spin polarization (both amplitude and orientation)
of the outgoing current could be controllable by electric means.\cite{spin
filtering diode,voltage1,rectification1,rectification2,voltage2,direction}
As an example, the concept of spin filtering diode\cite{spin filtering diode}
has been put forward based on the giant Zeeman splitting in semimagnetic
semiconductor heterostructures. Its salient feature lies in the large
difference of spin polarization when switching the polarity of the dc bias
applied to the device. The rectification of spin current has also been
predicted in asymmetric systems composed of either a molecular wire\cite%
{rectification1} or a quantum dot\cite{rectification2} sandwiched by a
ferromagnetic and a nonmagnetic lead. In these systems the ferromagnetic
lead is essential to generate a spin-polarized current.

The presence of spin-orbit interaction (SOI) in semiconductors
provides a way to design spintronic devices without need for a
magnetic element or an external magnetic field. Several devices
utilizing multiterminal electron waveguides have been proposed to
generate spin-polarized currents by means of the SOI
alone.\cite{SOC filter1,SOC filter2,SOC filter3,SOC filter4,SOC
filter5} For a two-terminal stub waveguide structure, we have
shown that the SOI-induced effective magnetic field can generate
both spin localization inside the stub and spin polarization in
the transmitted electron beam near structure-induced Fano
resonances.\cite{SOC filter6} We have also shown that the SOI
alone can not generate a spin-polarized transmitted electron beam
in a two-terminal waveguide when the output lead supports only one
orbital channel.\cite{Zhai-symmetry} Inspired by this fact, we
will show, in this work, that the spin transport properties of a
hornlike waveguide can be utilized to devise a spin current diode
without need for a ferromagnetic material or a magnetic field.

Our system is illustrated in the inset of Fig.~1(c), where a
two-dimensional electron gas (2DEG) in the $(x,y)$ plane is
restricted to a waveguide along the $x$ direction by a hard-wall
confinement potential $V_{c}(x,y)$. The 2DEG is contained in an
asymmetric quantum well so that the SOI arises mainly from the
interfacial electric field (the Rashba mechanism). The waveguide
consists of three parts. The left (right) part has a length
$L_{1}$ ($L_{3}$) and a uniform width $W_{L}$ ($W_{R}$), connected
to the left (right) lead with the same width. A finite difference
between the widths of the left and right parts of the waveguide
($W_{L}$ and $W_{R}$) is essential for the proposed device. Since
we are concerned only with spin-unpolarized injection, the two
connecting leads are nonmagnetic and have a vanishing SOI. The
central part of the waveguide spans the region $(x_{0}$,
$x_{0}+L_{2})$ along the $x$ direction,
within which the waveguide width $W_{C}$ varies smoothly from the initial value $%
W_{L}$ to the final value $W_{R}$. To be specific, we take
\begin{equation}
W_{C}(x)=W_{L}+(W_{R}-W_{L})\sin ^{2}[\pi (x-x_{0})/(2L_{2})].
\end{equation}%
For simplicity we assume that the whole waveguide shares a common horizontal
central line (at $y=0$). The effective-mass Hamiltonian describing the
considered system reads
\begin{eqnarray}
H &=&\left[ \frac{1}{2m^{\ast }}(p_{x}^{2}+p_{y}^{2})+V_{c}(x,y)\right]
\sigma _{0}  \notag \\
&&+\frac{1}{2\hbar }\left[ \alpha (x,y)(\sigma _{x}p_{y}-\sigma _{y}p_{x})+%
\text{H.c.}\right] \text{.}  \label{Hamiltonian}
\end{eqnarray}%
Here $m^{\ast }$ is the effective mass of electrons, $p_{x}$ and $p_{y}$ are
the in-plane momentum components, $\sigma _{x},\sigma _{y}$, and $\sigma
_{z} $ are the three Pauli matrices, and $\sigma _{0}$ is the $2\times 2$
unit matrix. The Rashba SOI strength $\alpha (x,y)$ is assumed to be uniform
(with a value $\alpha $) in the central part of the waveguide and decreases
adiabatically down to zero in the transition regions of the entrance and
exit. We take the spin quantum axis to be along the transverse $y$
direction, so that $|\uparrow \rangle =(1,i)^{T}/\sqrt{2}$ and $|\downarrow
\rangle =(1,-i)^{T}/\sqrt{2}$ represent the spin-up and spin-down states,
respectively.

A real-space discretization of Eq.~(\ref{Hamiltonian}) yields a
tight-binding model, which can be solved numerically by means of the
recursive Green's function method\cite{book} to obtain the outgoing wave
amplitudes. The Landauer-B\"{u}ttiker formula is then used to determine the
spin-resolved conductances $G_{\sigma ^{\prime },\sigma }$ ($\sigma ^{\prime
},\sigma =\pm 1$ or $\uparrow \downarrow $), which depend both on the
incident spin states $|\sigma \rangle $ in one lead and on the outgoing spin
states $|\sigma ^{\prime }\rangle $ in the other lead. The transmitted spin
current in the linear-response regime is characterized by the spin
conductance $(G_{s;\mathbf{x}},G_{s;\mathbf{y}},G_{s;\mathbf{z}})$.\cite%
{Zhai-symmetry} Since our system is invariant under the operation $\hat{R}%
_{y}\hat{\sigma}_{y}$, where $\hat{R}_{y}$ is the reflection $y\rightarrow -y
$, the spin conductance could be nonvanishing only along the $y$ direction
and is given by
\begin{equation}
G_{s;\mathbf{y}}=\frac{-e}{4\pi }\frac{G_{\uparrow ,\uparrow }+G_{\uparrow
,\downarrow }-G_{\downarrow ,\downarrow }-G_{\downarrow ,\uparrow }}{e^{2}/h}%
\text{.}  \label{spin conductance}
\end{equation}%
In the calculations we have chosen to fix the size parameters $%
W_{L}=W_{R}/2=100$ nm and $L_{1}=L_{2}=L_{3}=100$ nm. The electron effective
mass has been taken to be $0.041$ $m_{0}$ ($m_{0}$ is the free-electron
mass), which is appropriate to an InGaAs quantum well system.\cite{Sato}

In Fig.~1 we plot the total charge conductance $G=G_{\uparrow ,\uparrow
}+G_{\uparrow ,\downarrow }+G_{\downarrow ,\downarrow }+G_{\downarrow
,\uparrow }$ and the normalized spin conductance $G_{s;\mathbf{y}}$ [in unit
of $-e/(4\pi )$] as functions of the Fermi wave vector $k_{F}=(2m^{\ast
}E_{F}/\hbar ^{2})^{1/2}$, where $E_{F}$ is the electron Fermi energy, for
several values of the SOI strength $\alpha $. Here $k_{F}$ is given in units
of $k_{1}=\pi /W_{L}$, a value corresponding to the first subband energy $%
E_{1}$ in the narrow lead. The charge conductance exhibits a
steplike feature [see Fig.~1(a)] and is determined by the number
of propagating modes in the narrow lead, $N_{c}(E_{F})$. This
indicates a negligible backscattering when electrons traverse the
considered waveguide structure from the narrow lead to the wide
lead (the forward transport direction), that is
\begin{equation}
G_{\uparrow ,\downarrow }^{L\rightarrow R}+G_{\downarrow ,\downarrow
}^{L\rightarrow R}\approx N_{c}(E_{F})e^{2}/h\approx G_{\downarrow ,\uparrow
}^{L\rightarrow R}+G_{\uparrow ,\uparrow }^{L\rightarrow R}\text{.}
\label{conductance}
\end{equation}%
The left-to-right and right-to-left charge conductances,
$G^{L\rightarrow R}$ and $G^{R\rightarrow L}$, are identical due
to the time reversal symmetry. In contrast, the spin conductance
changes remarkably once the transport direction is reversed. Under
the forward bias, the spin conductance fluctuates around a single
plateau in the whole considered energy region [see Fig.~1(b)]. The
plateau moves up as the SOI strength increases. The derivation of
$G_{s;\mathbf{y}}^{L\rightarrow R}$ from the plateau value occurs
in the energy region $E_{1}<E_{F}<4E_{1}$ and near the onset of
subbands in the narrow lead. The spin polarization of the current
is the ratio between the normalized spin conductance and the
normalized charge conductance.\cite{Zhai-symmetry} As a result,
the spin polarization exhibits a steplike decrease as the Fermi
energy increases. When electrons are incident from the wide lead,
the spin conductance and thus the spin polarization is greatly
suppressed [see Fig.~1(c)]. A vanishing spin current is found in
the energy region of $E_{F}<4E_{1}$, in which the outgoing lead
(the left lead in this case) can support the lowest orbital mode
only. This is in full agreement with the prediction of
Ref.~\onlinecite{Zhai-symmetry}. When the outgoing lead supports
two or more propagating orbital modes, the spin conductance can be
finite but it is rather small in general [Fig.~1(c)]. A narrow
peak is observed near the onset of a subband (with subband index
$>1$) of
the outgoing lead, which is due to SOI-induced Fano resonance.\cite%
{Fano-Rashba}

The contrast in the spin conductance between the forward and
backward transport directions indicates a spin current diode even
in the small bias condition (the linear-response regime). The spin
current of the "on" state (the forward biased case) can be
controlled by the SOI strength. For the "off" state (the backward
biased case), the spin current is weak when the charge conductance
is on a quantized plateau or vanishing when the Fermi energy is in
the region of $[E_{1},4E_{1}]$. The results can be understood as
follows. The spin conductance comes from two parts
($G_{s;y}=G_{s1}+G_{s2}$). One is the difference between the two
spin-conserved conductances ($G_{s1}\propto
G_{\uparrow,\uparrow}-G_{\downarrow,\downarrow}$) and the other
one is the difference between the two spin-flipped conductances
($G_{s2}\propto G_{\uparrow,\downarrow}-G_{\downarrow,\uparrow}$).
We first consider the situation that electrons are incident from
the left (narrow) lead. From Eq.~(\ref{conductance}) we know that
the two parts, $G_{s1}$ and $G_{s2}$, are almost identical. Thus,
the spin conductance $G_{s;\mathbf{y}}^{L\rightarrow R}$ can be
expressed in terms of the difference between the two spin-flipped
conductances,
\begin{equation}
G_{s;\mathbf{y}}^{L\rightarrow R}\approx -e/(2\pi )(G_{\uparrow ,\downarrow
}^{L\rightarrow R}-G_{\downarrow ,\uparrow }^{L\rightarrow R})/(e^{2}/h)%
\text{.}
\end{equation}%
Such a difference is reflected by the variations of spin-flipped
transmissions $T_{q\bar{\sigma}\leftarrow p\sigma }^{L\rightarrow
R}$ shown in Fig.~2. Here, $\bar{\sigma}=-\sigma $, while $p$ and
$q$ are the indices of the incident and outgoing modes,
respectively. The $\hat{R}_{y}\hat{\sigma}_{y}$ symmetry of the
considered system implies
\begin{equation}
T_{q\bar{\sigma}\leftarrow p\sigma }^{L\rightarrow R}=0,\text{ }p-q\equiv 0%
\text{ }\text{mod 2.}
\end{equation}%
It can be seen that each nonvanishing transmission $T_{q\bar{\sigma}%
\leftarrow p\sigma }^{L\rightarrow R}$ is remarkable only within an energy
window. When the left lead supports only a single orbital channel
[Figs.~2(a) and 2(b)], the spin-flipped transmission for the spin-down
injection is much larger than that for the spin-up injection. This can be
explained by examining the SOI-induced mode mixing between subbands of
different spins in the central part of the considered waveguide.\cite%
{adiabatic} As two or more orbital modes are allowable for conducting in the
left lead ($p>1$), $T_{q\downarrow \leftarrow p\uparrow }^{L\rightarrow R}$
can be remarkable and even exceed the corresponding $T_{q\uparrow \leftarrow
p\downarrow }^{L\rightarrow R}$ in certain energy windows [see
Figs.~2(c)-2(f)]. However, the spin-flipped transmissions $T_{q\uparrow
\leftarrow p\downarrow }^{L\rightarrow R}$ for spin-down injections are seen
to be in general larger than their corresponding spin-flipped transmissions $%
T_{q\downarrow \leftarrow p\uparrow }^{L\rightarrow R}$ over large energy
regions. Furthermore, there exists such an outgoing channel $q=p^{\prime }$
that $T_{p^{\prime }\downarrow \leftarrow p\uparrow }^{L\rightarrow R}$ is
much smaller than $T_{p^{\prime }\uparrow \leftarrow p\downarrow
}^{L\rightarrow R}$. The combination of these facts gives rise to a nearly
constant spin conductance.

When the transport direction is reversed, the spin-resolved conductance can
be obtained from the relation imposed by the time reversal symmetry,
\begin{equation}
G_{\sigma ,\sigma }^{R\rightarrow L}=G_{\bar{\sigma},\bar{\sigma}%
}^{L\rightarrow R}\text{, }G_{\bar{\sigma},\sigma }^{R\rightarrow L}=G_{\bar{%
\sigma},\sigma }^{L\rightarrow R}\text{.}  \label{Time
reversal}
\end{equation}
This relation together with Eq.~(\ref{conductance}) indicates a cancellation of
$G_{s1}$ and $G_{s2}$ and thus results in $G_{s;\mathbf{%
y}}^{R\rightarrow L}\approx 0$, as observed in Fig.~1(c). From the above
analysis one can see that the spin current diode proposed here relies only
on two gradients: the quantized conductance and the difference of the two
spin-flipped conductances. Equation~(\ref{Time reversal}) also indicates
that for a spin-conserved system, such as the system studied in Ref.~%
\onlinecite{spin filtering diode}, the diode function of spin current can be
performed only in the nonlinear-response regime.

In conclusion, we have proposed a spin current diode based on a
waveguide connected to two leads with different width. It is
demonstrated that the spin conductance fluctuates around a
constant value in a wide range of incident energy when electrons
are incident from the narrow lead. When the transport direction is
reversed, the spin conductance is suppressed strongly. The
rectification of spin current is achievable even in the
linear-response regime and thus the proposed diode can work at a
low power consumption condition. The SOI alone is utilized to
realize such a function of spin current rectification.

F. Zhai was supported by the NSFC (Grant No. 10704013) and the training fund
of young teachers at Dalian University of Technology. K. Chang was supported
by the NSFC (Grant No. 60525405) and the knowledge innovation project of the
Chinese Academy of Sciences. H. Q. Xu acknowledges supports from the Swedish
Research Council (VR) and from the Swedish Foundation for Strategic Research
(SSF) through the Nanometer Structure Consortium at Lund University.

\newpage
\begin{center}
\textbf{Figure Captions}
\end{center}

\hfill

\begin{figure}[htbp]
\begin{center}
\includegraphics[height=8.5cm,width=8.5cm,bb=0 0 596
528]{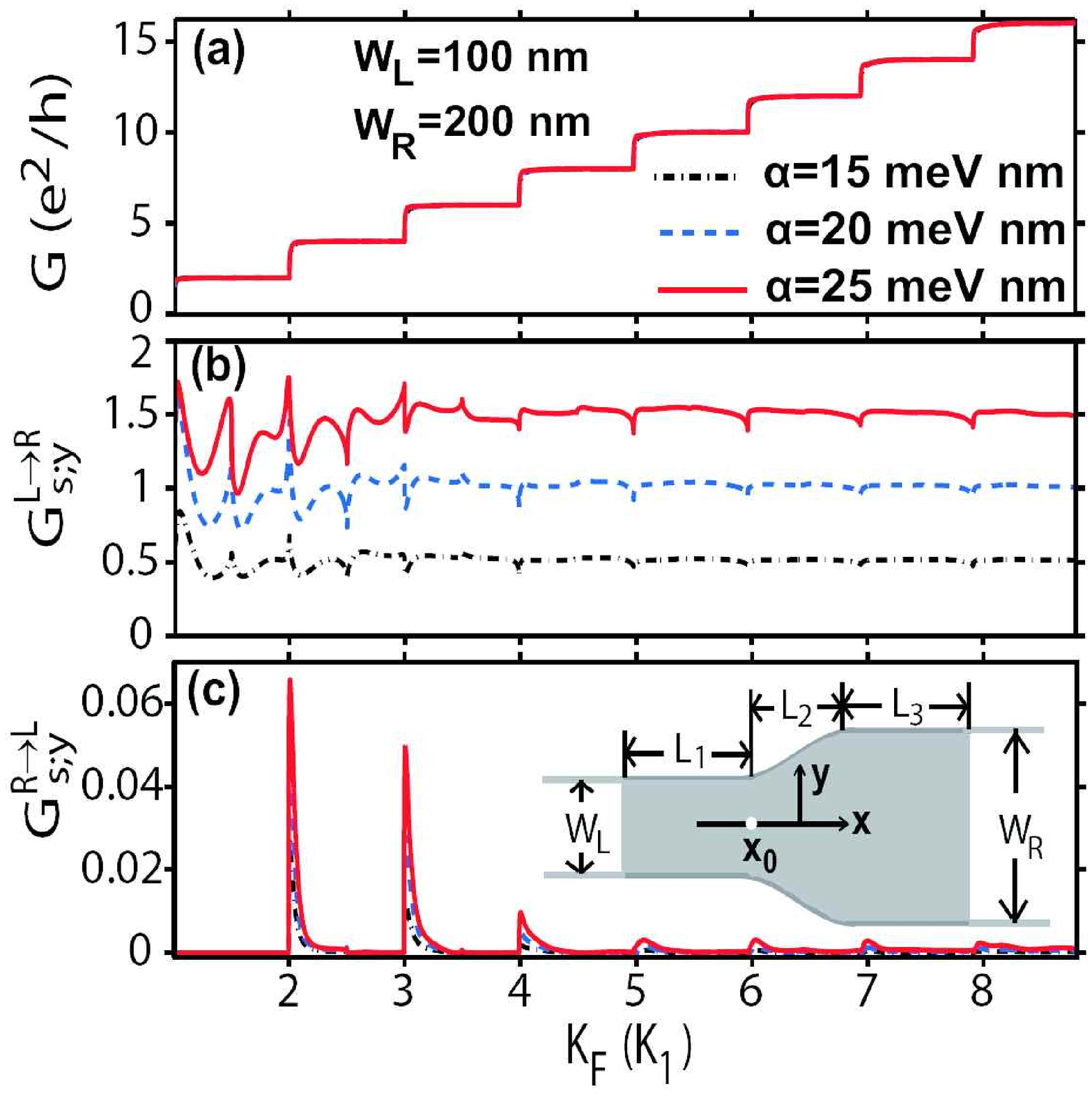}
\end{center}
\caption{(Color online) Conductance spectra of a two-terminal
horn-like waveguide structure with the Rashba SOI plotted as
functions of the Fermi wave vector for spin-unpolarized electron
injections: (a) total conductance $G$ (in unit of $e^2/h$); (b)
and (c) spin conductances $G_{s;y}^{L\rightarrow R}$ and
$G_{s;y}^{R\rightarrow L}$ [both in unit of $-e/(4\pi)$] for the
forward and backward transport directions, respectively. The inset
in panel (c) illustrates schematically the considered waveguide
structure. The structural parameters are given in the text.}
\label{fig:Fig1}
\end{figure}

\hfill

\begin{figure}[htbp]
\begin{center}
\includegraphics[height=8.5cm,width=8.5cm]{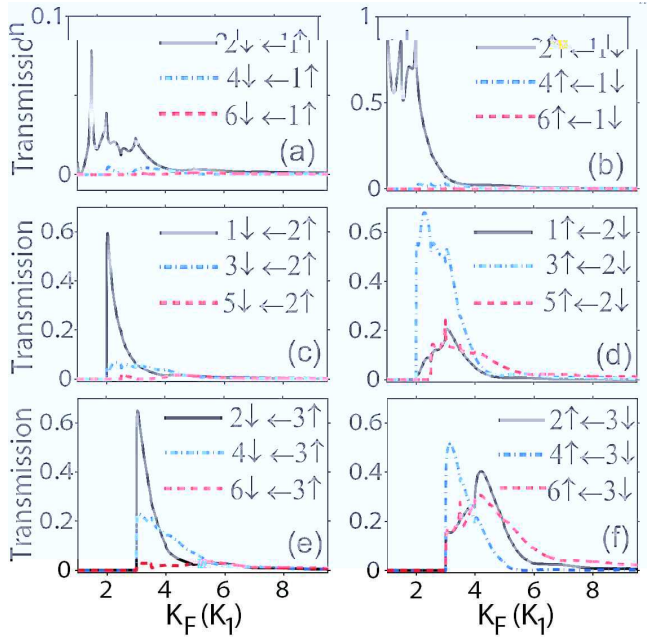}
\end{center}
\caption{(Color online) Typical spin-flipped transmission
probabilities $T_{q\bar{\sigma}\leftarrow p\sigma }^{L\rightarrow
R}$ for electrons with spin $\sigma$ incident from the subband $p$
in the narrow lead scattering into the subband $q$ in the right
lead with opposite spin. The structural parameters are the same as
those in Fig.~1 and the Rashba SOI strength is set at
$\protect\alpha = 25$ meV nm. Note that for the considered
structure depicted in the inset of Fig.~1(c),
$T_{q\bar{\sigma}\leftarrow p\sigma }^{L\rightarrow R}$ vanishes
when modes $p$ and $q$ have the same parity.} \label{fig:Fig2}
\end{figure}

\end{document}